\documentclass[letter,11pt]{article}
\pdfoutput=1 
\usepackage[utf8]{inputenc} 
\usepackage{jcappub} 
\newcommand{\be}{\begin{equation}}
\newcommand{\ee}{\end{equation}}
\newcommand{\beq}{\begin{equation}}
\newcommand{\eeq}{\end{equation}}
\newcommand{\bea}{\begin{eqnarray}}
\newcommand{\eea}{\end{eqnarray}}
\newcommand{\bear}{\begin{eqnarray}}
\newcommand{\eear}{\end{eqnarray}}
\newcommand{\mstar}{\ensuremath{M_\star}}
\newcommand{\rstar}{\ensuremath{R_\star}}

\usepackage{graphicx}  
\usepackage{dcolumn}   
\usepackage{bm,relsize}        
\usepackage{amssymb,amsmath,mathrsfs}
\usepackage{mathtools}
\usepackage{textcomp}
\usepackage{wasysym}
\usepackage{slashed}
\usepackage{multirow}
\usepackage{lipsum, color}
\usepackage[usenames,dvipsnames,svgnames]{xcolor}
\usepackage{booktabs}
\usepackage{xfrac}

\usepackage[capitalize]{cleveref}
\usepackage{subfig}

\newcommand{\eV}{\mathrm{eV}}
\newcommand{\mueV}{\mu\mathrm{eV}}

\newcommand{\Mhc}{M_{h,c}}
\newcommand{\Msc}{M_{\star,c}}

\newcommand{\Fig}[1]{Fig.~\ref{#1}}

\newcommand{\Eq}[1]{Eq.~\ref{#1}}

\newcommand{\ie}{{\it i.e.}}

\begin{document}
\begin{flushright}
\texttt{FERMILAB-PUB-24-0295-T}
\end{flushright}

\title{Axion Stars: Mass Functions and Constraints}

\author[a,b]{Jae Hyeok Chang,}
\author[a]{Patrick J.~Fox,}
\author[a,c]{and Huangyu Xiao}

\affiliation[a]{Theory Division, Fermi National Accelerator Laboratory, Batavia, IL 60510, USA}
\affiliation[b]{Department of Physics, University of Illinois Chicago, Chicago, IL 60607, USA}
\affiliation[c]{Kavli Institute for Cosmological Physics, University of Chicago, Chicago, IL 60637, USA}
\emailAdd{jhchang@fnal.gov}
\emailAdd{pjfox@fnal.gov}
\emailAdd{huangyu@fnal.gov}

\abstract{The QCD axion and axion-like particles, as leading dark matter candidates, can also have interesting implications for dark matter substructures if the Peccei-Quinn symmetry is broken after inflation. In such a scenario, axion perturbations on small scales will lead to the formation of axion miniclusters at matter-radiation equality, and subsequently the formation of axion stars. Such compact objects open new windows for indirect searches for axions. We compute the axion star mass function based on recent axion minicluster studies and Bose star simulations. Applying this mass function, we find post-inflation axion-like particles with masses $1.8 \times 10^{-21}~{\rm eV}<m_a< 3.3 \times 10^{-17}~{\rm eV}$ are constrained by the lack of dynamical heating of stars in ultrafaint dwarfs. We also find that current microlensing surveys are insensitive to QCD axion stars. While we focus on the gravitational detectability of axion stars, our result can be directly applied to other interesting signatures of axion stars, {\it e.g.} their decay to photons, that require as input the abundance, mass, and density distribution of axion stars.
}

\maketitle

\section{Introduction}

The QCD axion, currently a leading solution to the strong CP problem in quantum chromodynamics (QCD) that addresses the experimental nearly zero value of the neutron electric dipole moment, is also a viable dark matter candidate \cite{Peccei:1977hh,PhysRevLett.40.279,PhysRevLett.43.103,Abbott:1982af, Dine:1982ah,Preskill:1982cy, Peccei:2006as} and has gained significant attention in recent years, stimulating many experimental efforts on the direct detection of axion dark matter \cite{ADMX:2018gho,ADMX:2019uok,Alesini:2019ajt,Lee:2020cfj,Alesini:2022lnp}. In the post-inflationary scenario of axion, Peccei-Quinn symmetry is broken after inflation and axions develop different field values across different horizons, leading to a predictable relic abundance that is independent of initial conditions \cite{Buschmann:2021sdq,Gorghetto:2020qws}, which sets an important target for QCD axion experiments. For those axion masses, $ m_a\lesssim \rm meV$, the occupation number of axion dark matter is so high that it can be considered as wave-like dark matter which acts as a coherently-oscillating classical field.
On the other hand, the current spectrum of axion possibilities is broad, which motivates us to consider axion-like particles (ALPs) that do not solve the strong CP problem but have a broader range of axion parameters~\cite{Arvanitaki:2009fg}. 

The post-inflationary scenario of axions or axion-like particles involves interesting cosmological consequences such as the formation of axion miniclusters, also referred to as minihalos, and axion stars as early as matter-radiation equality \cite{Hogan:1988mp,Kolb:1993zz,Kolb:1995bu}, which opens a new window for indirect searches for axions that are complimentary to existing efforts on direct detection. 
The search for signatures of axion minihalos is not as mature as the search for the diffuse axion background.
Specifically, they have not imposed any constraints on the parameter space of QCD axions. This is partially due to the computational challenge of simulating the evolution of axion miniclusters but more because of the observational challenge of detecting them gravitationally. Even though axion miniclusters form at very early times, they are still too diffuse to be detectable with femto-, pico-, and microlensing surveys \cite{Kolb:1995bu,Fairbairn:2017sil,Katz:2018zrn}. There are a few future proposals that might eventually detect QCD axion miniclusters such as pulsar timing arrays \cite{Dror:2019twh,Ramani:2020hdo,Lee:2020wfn}, microlensing of highly magnified extragalactic stars \cite{Dai:2019lud}, and fast radio burst (FRB) timing \cite{Xiao:2024qay}. Axion stars, generally expected to form at the center of axion miniclusters, are more compact objects and can potentially increase sensitivity to the QCD axion or axion-like particles. 
This motivates us to study the properties of stars forming in these minihalos, specifically the timescale for their formation and the mass and density of the stars formed; for both the QCD axion and axion-like particles.  With this axion star mass function in hand, it is possible to determine the prospects for their detection through various astrophysical observables.

Enabled by recent progress in our understanding of the condensation of wave-like dark matter and the evolution of axion miniclusters, from both analytical and numerical studies, we can quantitatively study the mass distribution of axion stars and their compactness. There are still uncertainties that remain to be sorted out by future numerical studies on the exact growth model of axion stars or generic Bose stars. 
In the spirit of motivating indirect searches of axion stars as a powerful probe to axion dark matter in the post-inflationary scenario and stimulating new search strategies, we calculate the mass function of axion stars using a semi-analytical method for both QCD axion and axion-like particles, which can be further used to compare with other observations.

In this work, we focus on the gravitational interactions of axion stars, which are independent of the details of the axion self-couplings or couplings to SM fields.  One could also think of other observables such as radio signals, coming from the axion-photon coupling, which we leave for future work. Our proposal does not rely on the couplings between axions and standard model particles but only on the gravitational interaction. Note that only post-inflation axions may lead to the formation of axion miniclusters and axion stars. Meanwhile, the post-inflationary scenario reduces the axion diffuse flux that is the target of direct detection because most axions are bounded inside axion miniclusters or axion stars, which requires better sensitivity for direct detection \cite{Eggemeier:2022hqa,OHare:2023rtm}. Therefore, our work opens an exciting new window for indirect searches for the axion dark matter complementary to existing direct detection efforts.

This work is organized as follows: In Sec.~\ref{sec:overview}, a brief overview on the current status of axion minicluster and axion star study is presented. In Sec.~\ref{sec:axion_star_mf}, we perform our new calculation on the axion star mass function for post-inflation axion models. In Sec.~\ref{sec:observation}, we apply our axion star mass function to current observables and obtain new limits on axion masses. Our bound on the axion mass is $m_a>3.3\times 10^{-17}$ eV, which is currently the leading constraint for post-inflation axions. In Sec.~\ref{sec:discussion}, we present the conclusion and discussion of this work.

\section{Overview of Axion Stars and Axion Minihalos}\label{sec:overview}

Axion stars, gravitational bound states of axion dark matter, have attracted much interest and have been extensively studied in recent years. Generically, there are two types of compact objects formed from axions: axion miniclusters and axion stars. 
Axion miniclusters are gravitationally virialized objects formed from density perturbations on small scales, which naturally arise from isocurvature axion perturbations if the Peccei-Quinn symmetry is broken after inflation. The formation of axion miniclusters starts as early as the time of matter-radiation equality, after which the minihalos will subsequently merge and grow in mass as time evolves. In the literature, the term ``axion minihalo" has been used interchangeably with ``axion minicluster'', to emphasize the similarity to the usual dark matter halos, especially regarding the density profile \cite{Xiao:2021nkb}. We will refer to these virialized halo objects as axion minihalos in the rest of this work.

On the other hand, axion stars are more compact than axion minihalos. Axion stars are the ground state of axion field configurations sustained by a balance between gravitational attraction and the repulsion of gradient pressure; axion stars do not necessarily emit light like normal stars do. Therefore, axion stars are expected to form through Bose-Einstein condensation at the center of axion minihalos due to their large central density. There are two branches of axion star solutions: the dilute branch has a balance between gravity and axion kinetic pressure, and the dense branch has an unstable equilibrium between the attractive forces of gravity and axion self-interactions and the repulsion of kinetic pressure. Those two branches are separated by a critical mass, above which dilute axion stars will collapse into dense axion stars and subsequently explode by emitting relativistic axions. Such axion star explosions can continue to occur recurrently if axion accretion is fast enough, which are called recurrent axinovae, leading to the depletion of axion dark matter \cite{Fox:2023xgx}. In the scenario of QCD axion stars, however, one can show the axion star mass is one order of magnitude below the critical mass even considering the heaviest axion star formed in the current Universe \cite{Fox:2024xxx}. Therefore QCD axion stars are naturally stable dark compact objects supported by the axion kinetic pressure. For ALPs, axion stars can also be stabilized if their self-couplings are sufficiently small and do not trigger axinovae.

There are extensive studies on the stable solutions of axion stars (or generic Bose stars) \cite{Braaten:2015eeu,Schiappacasse:2017ham,Chavanis:2017loo,Visinelli:2017ooc,Guerra:2019srj,Delgado:2020udb,Zhang:2024bjo}, the formation \cite{Levkov:2018kau,Chavanis:2019faf,Chavanis:2020upb,Hertzberg:2020hsz}, and the collapse of axion stars \cite{Braaten:2016dlp,Levkov:2016rkk,Chavanis:2016dab,Zhang:2020bec}. However, the population analysis and detectable signatures are still lacking. Recently, there have been calculations to determine the axion star explosion rate \cite{Du:2023jxh,Fox:2023xgx}, the detection of axion star explosions \cite{Eby:2021ece,Arakawa:2023gyq,Arakawa:2024lqr,Eby:2024mhd}, and the encounter rate with axion minihalos and neutron stars \cite{Maseizik:2024qly}. The mass function and density of axion minihalos, as the environments for axion star formation, are crucial to determining the formation rate and population of axion stars. In this work, we adopt the state-of-art axion minihalo population and axion star formation models to systematically study the axion star mass function. In what follows, we will give an overview of the current literature on axion star solutions, the evolution of axion minihalos, and the formation of axion stars. Applying these theoretical tools, we will derive a mass function of axion stars, which will be important inputs for indirect searches of axion dark matter.

\subsection{Stable configurations of axion stars}

The stable axion-field configuration for the self-gravitating bound-state of non-relativistic axions can be found by solving the Schrödinger-Poisson equations (see \cite{Zhang:2018slz,Eby:2019ntd} for reviews). One could obtain generic solutions for the normal modes of axion stars (or Bose stars) by solving the wave function equations \cite{Chan:2023crj}. 
The ground state of a dense gas of axions is an axion star surrounded by a less dense gas of free axions.  Thus, one expects that, given enough time, these objects will form in dense environments.
The axion field or ALP field $a$, is a pseudo-Goldstone boson of a global $U(1)_{PQ}$ symmetry broken at a scale $f_a$. The axion mass term and self-couplings are generated through the instanton potential.
Since the $U(1)_{PQ}$ is anomalous under the color gauge group (or a QCD-like group) after the confinement scale, $\Lambda$, the axion potential will be generated through instanton effects, leading to a light mass term for axions. 

The axion potential takes the form \cite{GrillidiCortona:2015jxo}
\be
\label{eq:VaxionFull}
V(a) =- \frac{\Lambda^4}{c_{ud}}\sqrt{1-4c_{ud}\, \sin^2\left(\frac{a}{2f_a}\right)}~.
\ee
For the QCD axion, $\Lambda\approx 80 \,\rm MeV$ and $c_{ud}\approx m_u m_d/(m_u+m_d)^2 \approx 0.2$. For ALPs, the scale associated with the axion potential is a free parameter and $\Lambda$ can take on a wide range of values.  Even though the above expression specifically describes the QCD axion potential, we expect ALPs to have a similar sine potential and very similar phenomenology for the self-couplings. In what follows we will only be interested in the mass and quartic interactions contained in the full potential.  Expanding Eq.~\ref{eq:VaxionFull} to quartic order one can find the mass term and an attractive self-interaction term 
\be
V(a)=\frac{1}{2}m_a^2 a^2 - \frac{\lambda}{4!}a^4~.
\ee
For the QCD axion $m_a=\Lambda^2/f_a$ and $\lambda = (1-3c_{ud})m_a^2/f_a^2$, that is the mass and self-coupling are determined by the strong coupling scale of QCD ($\Lambda$) and the breaking scale of PQ symmetry ($f_a$).  For ALPs, where the $\Lambda$ is a free parameter $m_a$ and $f_a$ can be treated as independent, we still expect $\lambda\ll 1 $.

If we expand the axion potential to higher-order terms, the next term to appear is a repulsive self-coupling, at dimension-6. The analysis of the initial collapse of axion stars is always governed by the quartic term and the repulsive term will only become important during the further contraction of axion stars once they pass the point of instability.  These higher order terms have effects on the remnants of axion star explosions. 
In this work, we focus on the study of gravitational couplings of axions, which will dominate the axion star configuration and the formation in most cases. The self-couplings cause the instability of axion stars when they continuously accrete axion dark matter and reach the critical mass.  For the QCD-axion, this critical mass is not reached in the age of the universe.  For the case of an ALP it is always possible to change the PQ breaking scale, $f_a$, such that the self-couplings are too small to lead to critical star formation at the star masses we are interested in.

In the following, we sketch how to determine the stable axion field configurations.
In the most generic case, the Klein-Gordon equation of axion field in curved spacetime can be written as
\bea
\label{eq:a_KG}
\frac{1}{\sqrt{-g}} \partial_\mu(\sqrt{-g} g^{\mu \nu} \partial_\nu a) = - V'(a).
\eea 
It is often not required to work with the full theory of General Relativity.
The axion stars, especially the stable ones on the dilute branch are nonrelativistic objects as long as $f_a \ll M_{\rm pl}$. Therefore we can always simplify the calculation by studying the self-gravitating axion stars in the following metric
\bea
\label{eq:metric}
ds^2= g_{\mu \nu}dx^\mu dx^\nu = (1+2\Phi)dt^2-(1-2\Phi)dr^2 -r^2d\Omega^2,
\eea
where $\Phi$ is the gravitational potential given by the self-gravity of axion stars and $d\Omega$ is the differential solid angle. In the non-relativistic limit, we can also use single harmonics to approximate the axion field, which gives the following expression
\bea
\label{eq:axion_profile}
a(r,t) = f_a \,\Theta(r)\cos(\omega_a t),
\eea
where $\Theta(r)$ is the spatial-dependent part of the axion field and $ \omega_a$ is the energy of the axions in the star.  For the non-relativistic stars $\omega_a= m_a + p^2/2m_a \approx m_a$. 
If we neglect the axion self-interaction and higher order terms, Eq.~\ref{eq:a_KG} becomes
\bea
\label{eq:a_schrodinger}
\pmb{\nabla}^2 \Theta = 2 \left(m^2_a \Phi+\frac{m^2_a -\omega^2_a}{2} \right) \Theta,
\eea
where the spatial Laplace operator can be written as
$\pmb{\nabla}^2 = \frac{\partial^2}{\partial r^2} + \frac{2}{r} \frac{\partial }{\partial r}$
since we study the ground state configuration of axions that is spherically symmetric. 
In the Newtonian limit, the gravitational potential can be simply determined from the energy density via the Poisson equation 
\bea
\label{eq:Poisson}
\pmb{\nabla}^2 \Phi = 4 \pi G \rho_\star,
\eea
with $\rho_\star$ being the density of axion stars that contains the rest mass, kinetic energy, gradient energy, and self-energy, which takes the form
\bea
\label{eq:rhostar}
\rho_\star = 
\frac{m_a^2 f_a^2 \Theta^2}{2} + \frac{f_a^2}{4}  \left(\frac{d \Theta}{dr}\right)^2  - \frac{\lambda f_a^4 \Theta^4}{64}.
\eea
Where we have taken the time average of the axion field. Without fully solving these equations numerically, one could alternatively use a Gaussian ansatz to approximate the field profile \cite{Chavanis:2011zi,Chavanis:2011zm,Chavanis:2016dab}, which can demonstrate the competing effects driving the physics of stabilizing the axion stars \cite{Visinelli:2017ooc}. 
With this method, the energy component of an axion star of mass $\mstar$ and radius $\rstar$ can be determined quantitatively as 
\be\label{eq:Estar}
E_\star = -\frac{G_N \mstar^2}{\rstar} + c_1 \frac{\mstar}{2\,m_a^2 \rstar^2} - c_2 \frac{\lambda\mstar^2}{12\,m_a^4 \rstar^3}~.
\ee
The sum of these terms gives the total energy, with individual components corresponding to the self-energy term from gravity, the gradient energy, and the self-energy from the axion potential.  The numerical coefficients, $c_i$, are solved numerically for the full system \cite{Ruffini:1969qy,Membrado:1989ke,Visinelli:2017ooc} to be $c_1= 9.9$, $c_2=0.85$. 
As expected, the formalism of axion star energy shows that the axion star is balanced by gravity and kinetic pressure (or gradient energy). By adding mass to the axion star, one increases the gravitational self-interaction, which requires more kinetic pressure to balance the gravity, leading to a smaller axion star radius since the gradient energy goes like $1/R_\star^2$ while gravitational energy goes like $1/R_\star$. However, such balance will eventually be broken down by the quartic self-coupling, with an energy component that scales as $1/R_\star^3$. At the critical point, the density of axion stars is so high that self-couplings become important, and kinetic pressure cannot balance it by further decreasing the radius. The mass-radius relation for axion stars can be found by minimizing $E_\star$, which corresponds to the relaxation to the ground state of axion configurations and gives the following radius
\be
\label{eq:rstarpm}
\rstar^\pm=\frac{c_1}{2\,G_N \mstar m_a^2}\left(1\pm \sqrt{1-\frac{c_2}{c_1^2}  \lambda G_N\mstar^2}\right)~.
\ee
The $\rstar^+$ root corresponds to the dilute branch of axion stars, which is the axion field configuration stabilized by kinetic pressure and gravity while the self-interaction is negligible. Since the quartic self-couplings can be ignored in the dilute branch, the term related to $\lambda$ can be dropped, leading to a mass-radius relation
\be
\label{eq:rstarp}
\rstar^+ = 9.9\frac{M_{\mathrm{pl}}^2}{m_a^2\mstar}\approx 2.6\times 10^5 \,{\rm km} \left(\frac{10^{-6}\rm eV}{m_a}\right)^2\left(\frac{10^{-12}M_\odot}{M_\star}\right)~.
\ee
In general, a lighter axion mass $m_a$ corresponds to a larger axion star due to the large de Broglie wavelength. On the other hand, a heavier axion star has a smaller radius, because it requires more kinetic pressure to balance the gravity.  Overall, from Eq.~\ref{eq:rstarp} we see that the density of a star grows as the fourth power of its mass, $\rho_\star\sim M_\star^4(G_N m_a^2)^3$.
This will have important implications when we study the gravitational detectability of QCD axion stars. Since more compact objects can be probed with observations such as microlensing, one needs to form a sufficiently heavy axion star to reach a high density for detection.
Therefore, the evolution and growth of axion stars will be a crucial part of the story  (Unfortunately, as we will discuss later in Fig.~\ref{fig:MRcontour}, the QCD axion stars will not grow to a mass that will be compact enough for the current sensitivity of microlensing).
As the mass of the axion star grows, Eq.~\ref{eq:rstarpm} shows that at some point a sensible solution can no longer be obtained.  This is due to the breakdown of the balance between gravity and gradient pressure caused by the non-negligible addition of attractive self-couplings, leading to the collapse of axion stars.
The maximal mass that still corresponds to a stabilized axion star is \cite{Chavanis:2011zi,Chavanis:2011zm,Chavanis:2022fvh}
\be\label{eq:mstarmax}
\mstar^\textrm{max} = \frac{10.7}{\sqrt{\lambda}}M_{\mathrm{pl}}\approx 1.16\times 10^{-10}M_{\odot}\left(\frac{\rm 10^{-6}eV}{m_a}\right)\left(\frac{f_a}{10^{12}\rm GeV}\right)~.
\ee
The two solutions (\ref{eq:rstarpm}) merge at this critical mass (The $\rstar^-$ root is sometimes called the critical branch, which is metastable and not relevant to our study). After the collapse of critical mass stars, we will obtain the dense branch of axion stars
which will soon emit relativistic axions \cite{Eby:2015hyx,Eby:2016cnq,Levkov:2016rkk} and are thus cosmologically unstable, with a lifetime \mbox{$\sim 10^3 m_a^{-1}$} \cite{Visinelli:2017ooc,Zhang:2020bec}. In this work, we will not consider strong self-couplings that will destabilize axion stars. This corresponds to a large $f_a$ that is not excluded by Ref.~\cite{Fox:2023xgx,Fox:2024xxx} that considers the collapse of axion stars and their observational constraints. In this work, we will focus on the stable axion stars and study their gravitational detectabilities.

\subsection{An overview on axion minihalos}\label{sec:axionminihalos}

Axion minihalos are natural products of axion cosmology in the post-inflationary scenario where Peccei-Quinn symmetry is broken after inflation \cite{Hogan:1988mp,Kolb:1993zz,Kolb:1995bu}. The axion field will randomly take different values at different horizon patches when its potential is flat. During the QCD phase transition, axions acquire their mass, converting the inhomogeneity of field values to matter density fluctuations. This leads to order unity matter perturbations at the scale corresponding to the size of horizon patches when axions obtain their mass.  These large perturbations subsequently and form axion minihalos at, or shortly after, matter-radiation equality. The axion isocurvature fluctuations at superhorizon scales follow a white noise power spectrum because different horizon patches were not in casual contact so there is no correlation.  Furthermore, because many horizon patches contribute to perturbations at superhorizon scales, the spectrum of fluctuations is Gaussian and the power spectrum fully describes the fluctuations. Note that these superhorizon modes will enter the horizon before matter-radiation equality and grow linearly with the scale factor.
From the white-noise power spectrum, one can study the population of axion minihalos at different cosmic times. 

The cosmological evolution of axion minihalos after its initial formation has been studied using semi-analytic techniques \cite{Enander:2017ogx,Blinov:2019jqc,Fairbairn:2017sil,Ellis:2020gtq,Ellis:2022grh} as well as full numerical simulations \cite{Zurek:2006sy,Eggemeier:2019khm,Xiao:2021nkb,Eggemeier:2022hqa,Shen:2022ltx}, which suggest $\sim75\%$ of the axion dark matter will be bound into axion minihalos in the late Universe due to the large axion perturbations. Even though they are abundant, it is very challenging to observe these axion minihalos. 
Some ideas might involve probing these structures via their photon signals such as the stimulated decay of axions \cite{Tkachev:2014dpa,Iwazaki:2014wka,vanWaerbeke:2018nyj,Dietrich:2018jov,Sun:2020gem,Buckley:2020fmh,Caputo:2018vmy,Sun:2021oqp,Dev:2023ijb,Sun:2023gic}, and axion-photon conversion in neutron stars \cite{Safdi:2018oeu,Battye:2019aco, Foster:2020pgt,Hook:2018iia}. Other proposals considered the lensing or photon signatures from axion stars \cite{Bai:2017feq, Hertzberg:2018zte,Hertzberg:2020dbk, Prabhu:2020pzm,Prabhu:2020yif,Carenza:2019vzg,Fujikura:2021omw,Escudero:2023vgv,Gan:2023swl,Yin:2024xov}. Our work will focus on the calculation of the axion star mass function and its gravitational effects, which will have important implications for the rate of detectable axion star events.

According to simulations of string dynamics \cite{Vaquero:2018tib, Buschmann:2019icd,OHare:2021zrq}, the dimensionless matter power spectrum $\Delta^2(k)$, after the axion string network has collapsed, follows an approximate (white noise) power law at large scales before becoming approximately constant at short distance scales and then ultimately decreasing at very small scales.  The exact form of the power spectrum is still under active study.  Here we choose to concentrate on the power law part of the power spectrum and to remain agnostic to the exact form once this power law breaks.  We do this by modeling the spectrum as a pure white noise power law at all scales but place a lower bound on halo masses so that we do not probe the non-white noise region of the power spectrum.  Thus, the dimensionless matter power spectrum $\Delta^2(k)$ takes the form
\begin{equation}\label{eq:power}
    \Delta^2(k)\equiv \frac{k^3}{2\pi^2}P(k)= A_0 \left(\frac{k}{k_{0}}\right)^3 \, ,
\end{equation}
where $k_0$, which is the comoving wavenumber where the matter power spectrum in full simulations becomes constant, and $A_0$ is the amplitude at this scale, which is also the peak amplitude of $\Delta^2(k)$.  Here $P(k)\equiv V^{-1}|\tilde{\delta}(\mathbf{k})|^2$ is the matter power spectrum, $V$ is the volume, $\tilde{\delta}$ is the Fourier transform of the linear dark matter overdensity.  For the QCD axion $A_0\approx 0.4$, while for ALPs $A_0\approx 0.3$ \cite{OHare:2021zrq}.  The scale where the power law breaks, $k_0$, is related to the comoving horizon scale at the time of axion oscillation, $k_{\rm osc} \equiv  a_{\rm osc}{ H}_{\rm osc}$.  For the case of the QCD axion $k_0\approx 3 k_{\rm osc}$, while for ALPs $k_0\approx k_{\rm osc}$.  Associated with the scale $k_0$ is a mass, which we denote the characteristic halo mass,
\begin{equation}\label{eq:Mhc}
\Mhc = \frac{4\pi}{3}A_{\rm 0}\left(\frac{1}{k_{\rm 0}}\right)^3  \bar{\rho}_{a,0}~.
\end{equation}
We focus only on halos whose mass is greater than $\Mhc$.  Halos with $M_h<\Mhc$ will correspond to a higher wavenumber, $k>k_0$, where the spectrum of perturbation is less well understood. 
These high wavenumbers do not contribute much mass to the total minihalo distribution.

The mass function of axion minihalos, especially at the high mass end, which constitutes the majority of the mass, is well studied with N-body simulations that also agree with a calibrated analytical formula \cite{Xiao:2021nkb}. A semi-analytical model by Press $\&$ Schechter \cite{1974ApJ...187..425P} gives the following halo mass function: 
\begin{equation}
\frac{df_h}{d \log M_h} = \sqrt{\frac{\nu}{2\pi}}{\rm exp}(-\nu/2)\frac{d \log\nu}{d \log M_h} \, ,
\label{eq:current_mf}
\end{equation}
where $M_h$ is the axion minicluster(or minihalo) mass and $df_h/d \log M_h$ is the mass fraction of axion minihalos per logarithmic mass \ie\ $f_h$ is the mass fraction of dark matter bound in minihalos. 
We choose  $df_h/d \log M_h$ to present the mass distribution of the population of axion minihalos, as it can characterize the most significant mass ranges of these objects that contain the majority of the total mass when considering logarithmic mass bins.
$f_h =M_h n_h/\bar{\rho}_{a,0}$ is the mass density fraction of axion in minihalos. $\nu$ is defined as:
\begin{equation}
\nu \equiv \frac{\delta_c^2}{\sigma^2(M_h)D(z)^2} \, ,
\end{equation}
where $\delta_c=1.686$ is the critical density required for spherical collapse and $D(z)$ is the growth function which is normalized to be $D(z_{i})=1$ where $z_i$ is the initial redshift in radiation dominated era.
For subhorizon modes, the growth function is described by the M$\rm\Acute{e}$sz$\rm\Acute{a}$ros equation which gives a solution 
\be
D(z)= 1+\frac{3}{2}\left(\frac{1+z_{\rm eq}}{1+z}\right)~. 
\ee
We are ignoring the effects of dark energy since axion star formation occurs early when dark energy is a subdominant fraction of the universe.
The variance of the axion perturbations, $\sigma^2(M_h)$, in the initial density fluctuations with a tophat filter of scale $R=(3M_h/4\pi\bar{\rho})^{1/3}$, can be calculated as: 
\begin{equation}
\sigma^2(M_h)\equiv \int \frac{dk}{k}\frac{k^3P(k)}{2\pi^2}\left \vert \widetilde{W}(kR)\right \vert^2,
\end{equation}
where $\widetilde{W}(x) = (3/x^3)[{\rm sin}(x)-x{\rm cos}(x)]$ 
is the spherical top-hat window function. There are other reasonable choices for a window function, {\it e.g.}\ sharp $k$-space filter, which give qualitatively similar results.
The variance of the white-noise power spectrum from axion perturbations is 
\begin{equation}\label{eq:sigma_M}
\sigma(M_h)=\sqrt{\frac{3 \pi}{2}\frac{\Mhc}{M_h}} \, .
\end{equation}
As we see in the above expression, the only relevant parameter for $\sigma(M_h)$ is the dimensionless quantity $x_h \equiv M_h/M_{h,c}$, which will give the full description of the dimensionless mass function in Eq.~\ref{eq:current_mf}.
At $\nu<1$, the exponential part of the mass function can be ignored and the mass function exhibits a power law $df_h/d \log M_h\propto x_h^{1/2}$. As discussed before, we cut off the mass ranges $M_h<M_{h,c}$ (or $x_h<1$) to filter out the contributions from high wave numbers; we only consider $x_h\ge 1$.

Therefore the mass distribution of axion minihalos is peaked at larger masses, which suggests that the observable is sensitive to the axion perturbations at relatively large scales. Note that the above mass function includes the white-noise power spectrum only, which only applies at $z\gtrsim 20$ when large-scale structures have not formed yet. Afterward, the infall of axion minihalos into large-scale structures will stop the evolution of the mass function, as extensively discussed in Ref.~\cite{Xiao:2021nkb}. In this work, we focus on studying axion minihalos around matter-radiation equality, which maximizes the axion star formation rate. Therefore, the above mass function that describes the minihalos at early times should suffice for our purpose.

The density of axion minihalos is also needed to compute the formation timescale of axion stars. Axion minihalos are highly concentrated substructures, with a Navarro-Frenk-White (NFW) \cite{Navarro_1996} density profile as tested in N-body simulations \cite{Xiao:2021nkb}. The size and density of an NFW halo can be characterized by the scale radius and the scale density, which will ultimately determine the axion star formation rate in axion minihalos.
As computed in N-body simulations \cite{Xiao:2021nkb}, the scale radius of axion minihalos formed from a white-noise power spectrum is given by:
\begin{equation}\label{eq:rs}
	r_s \approx 3125{\rm AU}\left(\frac{\Mhc}{10^{-12}M_{\odot}}\right)^{-1/2}
	\left(\frac{M_h}{10^{-6}M_{\odot}}\right)^{5/6},
\end{equation}
where $M_h$ is the axion minihalo mass. Given the scale radius and the minihalo mass, one can determine $\rho_s$ by:
\begin{equation}\label{eq:rhos}
	\rho_s=\frac{M_h}{4\pi r_s^3(\log (1+c)-c/(1+c))} \, ,
\end{equation}
where $c$ is the halo concentration number, which is defined by the ratio of virial radius to scale radius, $c=r_{\rm vir}/r_s$. Simulations show that the concentration of axion minihalos can be expressed as $c\approx 1.4\times 10^4\, (1+z)^{-1} \sqrt{M_{h,c}/M_h}$ \cite{Xiao:2021nkb}, which agrees well with an analytical model that the concentration of axion minihalos is roughly 4 at birth ($\sigma(M_h)=\delta_c$). At matter-radiation equality, most axion minihalos born at that time should have a concentration close to 4.
When we calculate the condensation timescale of axion stars in realistic axion minihalos later, the central density and velocity will be needed. For conservative estimates, we can use the scale density as a benchmark for the central density. 
The characteristic velocity can be estimated as the circular velocity at scale radius 
\begin{equation}\label{eq:halo_v_rho}
	v_{\rm c} = \sqrt{2\pi G r_s^2\rho_s \left(\log 4-1\right)}~.
\end{equation}

\subsection{Formation of axion stars in axion minihalos}
Since the axion star configuration, as discussed earlier in this section, is an energetically favored state of axion fields, the axion waves prefer to reach the axion star configuration as long as there is enough time for relaxation. Such processes can be understood as Bose-Einstein condensation of Bosonic dark matter.
Therefore, the relevant timescale of axion star formation is the scattering timescale between axions as wave-like particles, which approximately gives the condensation timescale of axion stars in simulations of axion fields \cite{Levkov:2018kau,Eggemeier:2019jsu,Chen:2021oot,Kirkpatrick:2020fwd,Jain:2023tsr} 
\be\label{eq:generaltimescale}
\tau  \sim \left(f_{\mathrm{BE}} n \sigma v\right)^{-1} \, ,
\ee
where $f_{\mathrm{BE}}=6\pi^2 n(m_a v)^{-3}$ is the phase density of axions, with $n$ the typical axion number density and $v$ is the axion speed.  The axion wave scattering cross-section is $\sigma$, this scattering can be due to axion self-couplings or gravitational interactions.

This scattering of axion waves is Bose-enhanced due to the large phase space density. Therefore, a light axion mass or a small velocity will greatly enhance the formation rate of axion stars. Quartic self-interactions of axions can further enhance the rate but also lead to the collapse of axion stars and interesting constraints on the attractive self-couplings \cite{Fox:2023xgx}.
In this work, we focus on the gravitational interactions of axions, which gives the minimal rate of axion star formation. All axion stars formed with negligible self-couplings will be cosmologically stable objects.
The cross-section of gravitational scattering is \mbox{$\sigma_{\mathrm{gr}}=8\pi (G_N m_a v^{-2})^2 \log(m_a v r_{\rm vir})$}, where the Coulomb logarithm applies a cutoff at the size of the axion minihalo, $r_{\rm vir}$.
Therefore, the gravitational condensation timescale is
\be
\label{eq:taugrav}
\tau = \frac{b}{48\pi^3}\frac{m_a v^6}{G_N^2 n^2\log \left(m_a v r_{\rm vir}\right)} \, .
\ee
The parameter $b\sim\mathcal{O}(1)$ is a numerical coefficient determined by numerical simulations \cite{Chen:2020cef}. $r_{\rm vir}= c \, r_s$ is the virial radius of the minihalo that provides the cutoff scale. We focus on the axion star formation around matter-radiation equality, and we choose $c=4$ similar to the choice in the previous subsection for a simple estimation of the logarithmic factor.
These formulae can calculate the relaxation scale of axion gas with a given velocity distribution and agree well with the results found in numerical simulations \cite{Levkov:2018kau,Eggemeier:2019jsu,Chen:2021oot,Chan:2022bkz}.

The gravitational condensation time, which is proportional to $v^6/\rho^2$, is dramatically shorter at a smaller radius in the halo center. In this work, we calculate the characteristic density and speed at the scale radius for conservative estimations.
Since the star formation rate peaks at the center, we generically expect the axion star to form at the halo center. While multiple stars may form in a halo when the self-interaction is strong enough and axion gas fragments to stars in different regions of one minihalo, we do not consider that possibility and focus on the scenario of only one axion star per axion minihalo. The story of the axion star formation rate is not yet finished since we have not determined the relevant mass scales of axion stars formed after the relaxation. 
It was recently found the growth of Bose stars in a halo environment is self-similar \cite{Dmitriev:2023ipv}, and the growth behavior can be described by 
\begin{equation}\label{eq:star_growth}
	\frac{(1+x_{\star}^3/\epsilon^2)^3}{(1-x_{\star})^5}\approx \frac{t+0.1\tau}{1.1\tau},
\end{equation}
where we define $x_{\star}(t) \equiv M_{\star}(t)/M_h$ with the axion star mass $M_\star$ and $x_{\star}=0$ at $t=\tau$, with $\tau$ the timescale for the minihalo system to become thermalized. The quantity $\epsilon \approx 3\Msc/M_h\approx10vM_{\rm pl}^2/(m_a M_h)$ is a combination of total mass and energy in the minihalo system, which is a measure of the star-halo mass relation. 
$\Msc$ is the characteristic mass scale of the axion star and can be well approximated by equating the virial velocity of the axion star and the halo velocity, which gives \cite{Levkov:2018kau,Eggemeier:2019jsu}
\begin{equation}
\label{eq:M_star_c}
    \Msc  \approx \frac{3 \,v}{G_N m_a}\approx 3\rho_a^{1/6} G_N^{-1/2}m_a^{-1} M_h^{1/3}~,
\end{equation}
where $v$ is the velocity of axion waves in axion minihalos, $\rho_a$ is the scale density of axion minihalos, and $M_h$ is the axion minihalo mass.
Sometimes the characteristic mass of axion stars $\Msc$ is also called the ``saturation" mass since the growth slows down after this mass is reached according to numerical simulations. However, the mass growth will continue even after $x_{\star}\sim\epsilon$ in this self-similar model discussed above. The power law is indeed different once $x_{\star}^3/\epsilon^2\gg 1$ is satisfied. As $x_{\star}$ approaches 1, it takes more and more time for the star growth, leading to the slow down of axion star formation. The above expressions can be used to calculate the mass of the axion stars once we have the axion minihalo properties.

\section{Axion Star Mass Functions}\label{sec:axion_star_mf}

While the formation of axion minihalos, or the formation of axion stars within certain backgrounds, has been studied, the current time mass distribution of axion stars is not well-known. Determining an exact axion star mass function requires extensive numerical study taking into account physics on widely separated scales and wave effects, but we provide an analytical estimate of the axion star mass distribution using several conservative assumptions. As a simple estimate, we suggest taking the axion minihalo distribution at the time of matter-radiation equality ($z\approx 3400$) and considering axion stars formed within each minihalo over one Hubble time from the time of equality. 

The first minihalo starts to form roughly at matter-radiation equality because the Hubble time at equality is approximately the same as the minihalo dynamical timescale needed to form a virialized minihalo \cite{Penston:1969yy,Chang:2018bgx}.
After the equality, axion stars can form and grow, but axion minihalos also grow and merge. As axion minihalos become heavier, the axion star formation time increases faster than the Hubble time, causing the growth of axion stars to cease. Also, we only consider one axion star per minihalo because they typically form in the center of minihalos, and if more than one stars are formed they quickly merge. Thus, we focus on the formation around matter-radiation equality, and the late evolution of axion stars such as mergers, which requires more extensive numerical studies, is not considered in this work. Our calculation underestimates the mass of individual stars, but we expect the total mass fraction in axion stars is well captured.
This simple estimate already suggests that an $ \mathcal{O}(1) $ fraction of axion dark matter forms axion stars, which can lead to interesting observational constraints and potential signals. 

Once we have the axion minihalo properties, the axion star mass formed within the minihalo can be determined.
We can calculate the axion star mass for a given axion mass $m_a$ and minihalo mass $M_h$, at a particular time, by solving \Eq{eq:star_growth}.  Which gives
\begin{equation}\label{eq:Ms}
M_\star(m_a,M_h) = M_h \, x_\star\left(\frac{1}{H(z_{\rm eq})}\right)~.
\end{equation}
Recall that $x_\star$ depends upon the typical density and size of the minihalo as well as the speeds of axions in the minihalo.  All of these are determined,  through Eqs.~\ref{eq:rs}, \ref{eq:rhos}, and \ref{eq:halo_v_rho}, by the minihalo mass (along with the parameters associated with the power spectrum, $k_0$ and $A_0$).
Since we limit ourselves to the case where there is at most one axion star per minihalo the axion star mass fraction is related to the axion minihalo mass fraction $df_\star/dM_h = (M_\star/M_h) df_h/dM_h$.  Furthermore, from the star-minihalo mass relation, the axion minihalo population described in \Eq{eq:current_mf} can be used to determine the axion star mass function by the following expression
\begin{equation}\label{eq:dfstar}
\frac{df_\star}{dM_\star} =  \frac{M_\star}{M_h}\frac{dM_h}{dM_\star} \frac{df_h}{dM_h} \, ,
\end{equation}
where $f_\star$ is the mass fraction of axions in axion stars to total ambient axions. We will show the mass distributions of axion stars for both QCD axions and ALPs, in the next two subsections.

\subsection{QCD Axion Stars}\label{sec:QCDaxionstars}

\begin{figure}[t]
\centering
\includegraphics[width=0.8\textwidth]{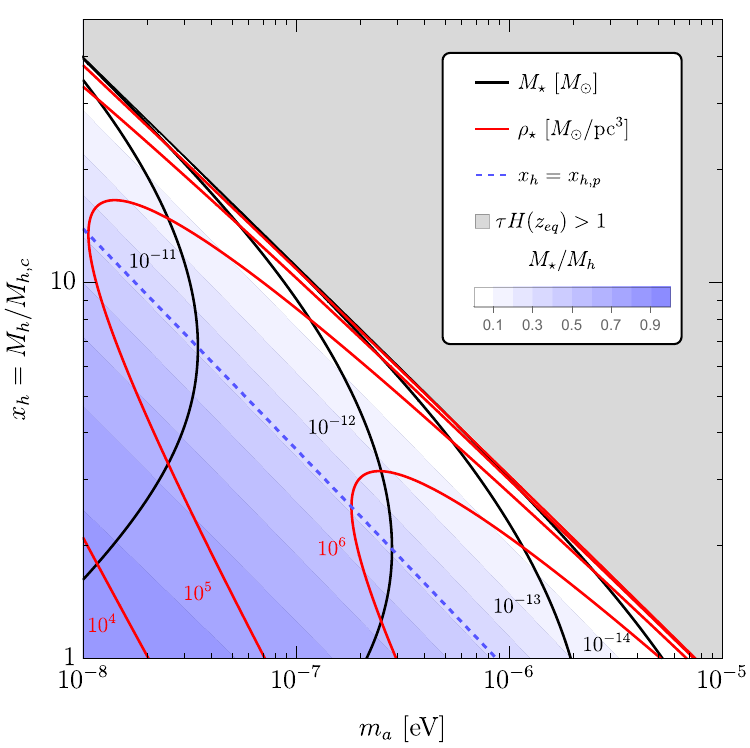}  
\caption{
This plot presents the axion star mass and density as a function of axion minihalo mass $M_h$ and axion mass $m_a$.  We scale the $y$-axis to use the mass ratio $x_h=M_h/M_{h,c}\ge 1$ to represent the minihalo mass. The black contours show different axion star masses ($10^{-14} M_\odot \lesssim M_\star \lesssim 10^{-11} M_\odot$), and the red contours show different densities of axion stars ($10^4 \, M_\odot/\textrm{pc}^{-3} \lesssim \rho_\star \lesssim 10^6 \, M_\odot/\textrm{pc}^{-3}$). The blue dashed line indicates the pivot minihalo mass $x_{h,p}$, which gives the maximum axion star mass for a given axion mass. The grey region does not lead to a sufficient axion star formation rate because the condensation timescale is too long compared to the Hubble time at matter-radiation equality. The ratio of the star to minihalo mass are shown as blue-shaded regions.
A smaller $x_h$ corresponds to a higher halo density and thus a larger axion star formation rate. Therefore, the axion star to halo mass ratio $M_\star/M_h$ increases with a smaller $x_h$. We focus on the mass range $m_a<10^{-5}\, \rm eV$ since we mainly study halos with $x_h>1$ that will dominate the population of axion minihalos. It is possible to have $x_h<1$, especially considering the effects from axion strings \cite{Gorghetto:2024vnp}, which will further enhance the axion star formation for larger $m_a$. 
}
\label{fig:MRcontour}
\end{figure}

\begin{figure}[t]
\centering
\includegraphics[width=0.95\textwidth]{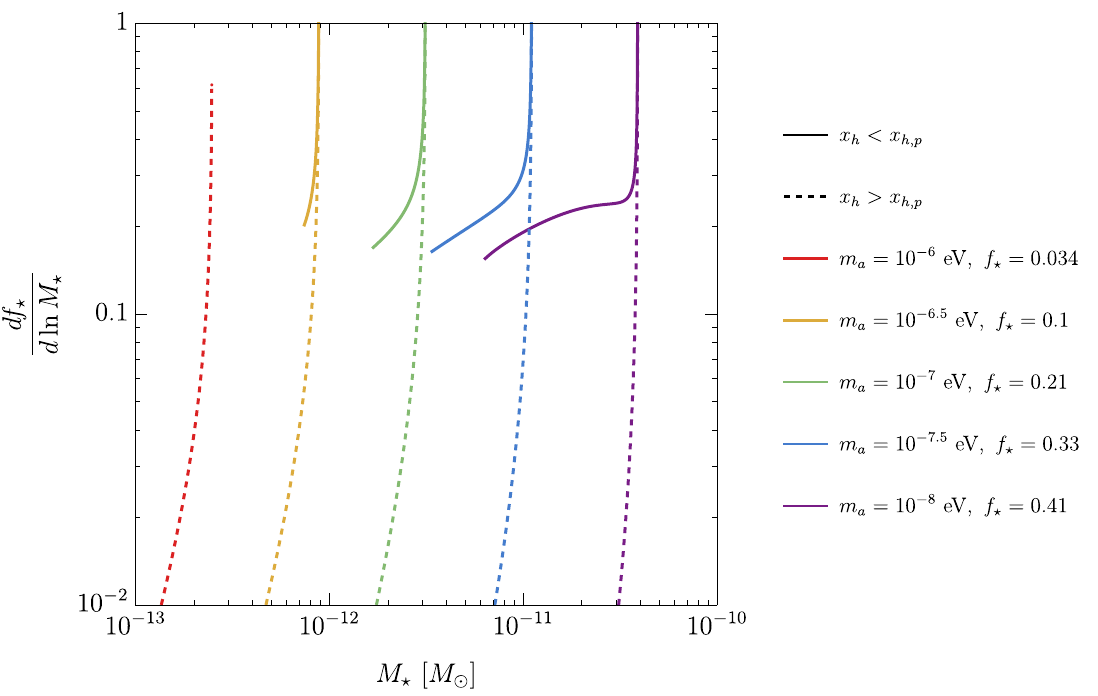}    
\caption{This plot presents the mass function of QCD axion stars for various axion masses $m_a$, which correspond to different typical values of axion star masses. The solid curves represent the contribution from $x_h<x_{h,p}$ while the dashed curves come from larger axion minihalos with $x_h>x_{h,p}$. Note that axion stars formed from smaller axion minihalos ($x_h<x_{h,p}$) will dominate the mass function. A smaller $m_a$ will lead to a larger dark matter fraction in axion stars, $f_\star$, due to the shorter gravitational condensation timescale.
}
\label{fig:massfunc}
\end{figure}

It is crucial to determine the QCD axion minihalo mass before obtaining the mass function of QCD axion stars, which will be discussed in this subsection. 
The time when axions start to oscillate is determined from $m_a(T_{\rm osc}) =1.6 H(T_{\rm osc})$, which will give the characteristic mass, $M_{h,c}$, of axion minihalos. 
Since the mass term of the QCD axion is induced by the temperature-dependent QCD instanton potential, it is suppressed at temperatures higher than the QCD confinement scale.
Therefore, the oscillation of the QCD axion is generally delayed compared to the naive expectation from its zero temperature mass. The temperature-dependent QCD axion mass is given by $m_a(T)\sim m_a(0) (T/\Lambda_{\rm QCD})^{-4}$ for $T\gtrsim\Lambda_{\rm QCD}$ \cite{Hertzberg:2008wr}. For the QCD axion mass range  we will consider, $m_a>10^{-8}$ eV, the oscillation temperature is higher than the QCD phase transition scale, $T_{\rm osc}\gtrsim \Lambda_{\rm QCD}$ and the temperature dependent mass must be considered. For smaller axion masses, oscillation will occur at the zero temperature mass.  We consider this possibility, over a broader range of masses in Sec.~\ref{sec:alpstars}.

The mass range for the QCD axion is motivated by the studies on the relic abundance, with lighter axion masses predicting a larger relic density since the oscillation starts later and the axion density is thus less diluted. The vacuum misalignment mechanism predicts a mass of $m_a\sim 10^{-5}$ eV for the correct relic density. Extra contributions to the axion density from string emissions will predict a relatively larger relic density and thus prefer larger masses \cite{Gorghetto:2018myk,Gorghetto:2020qws,Buschmann:2021sdq,Kawasaki:2018bzv,Saikawa:2024bta,Kim:2024wku}. However, non-standard thermal history such as an early-matter-dominated epoch that reheats after the QCD era can inject extra entropy into the Universe, diluting the axion density. We consider a broad QCD axion mass ranges $10^{-8}\, \eV < m_a < 10^{-5}\, \eV$ in this work. 

Using the full temperature-dependent mass of the QCD axion, the oscillation scale $k_{\rm osc}$ is \cite{Borsanyi:2016ksw,Vaquero:2018tib}
\begin{equation}
k_{\rm osc} \approx 14 ~ \textrm{pc}^{-1} \left(\frac{m_a}{\mueV} \right)^{0.17} \, .
\end{equation}
Given that $k_0 \sim 3k_{\rm osc}$ and $A_0 \sim 0.4$, as discussed earlier, the characteristic mass of axion minihalos \Eq{eq:Mhc} is
\begin{equation}
\Mhc \approx 7.0 \times 10^{-13} \, M_\odot \left(\frac{\mu \eV}{m_a} \right)^{0.51} \, .
\end{equation}
As for the halo mass function (see Sec.~\ref{sec:axionminihalos}) it is convenient to introduce the dimensionless parameter $x_h\equiv M_h/\Mhc>1$ to represent the axion minihalo mass in a $\Mhc$ independent way. Then combining \Eq{eq:rs}, \Eq{eq:rhos}, and \Eq{eq:M_star_c}, we can get the characteristic axion star mass
\beq
\Msc \approx 1.4 \times 10^{-13} \, M_\odot \, x_h^{1/12} \left(\frac{\mu \eV}{m_a} \right)^{1.17} \, .
\eeq
With the determination of the axion minihalo mass, we can obtain the formation timescale (\Eq{eq:taugrav}) for QCD axion stars:
\beq
\tau H(z_{\rm eq}) \approx \frac{0.083 \, x_h^{7/2} (m_a/\mueV)^{1.98}}{\log[29 \, x_h^{11/12}(m_a/\mueV)^{0.66}]}
\eeq
With these expressions, the QCD axion star mass present in a halo shortly after matter-radiation equality can be computed,
from \Eq{eq:Ms}.

In Fig.~\ref{fig:MRcontour}, we present the properties of QCD axion stars formed in a given axion minihalo with a mass of $M_h$ within the Hubble time at matter-radiation equality. We also show the contours of different masses and densities of axion stars as a function of $m_a$ and $M_h/M_{h,c}$, where $M_{h,c}$ is the characteristic mass of initial axion minihalos. 
For larger minihalos, the axion star mass decreases due to a more suppressed formation rate and the corresponding contribution to the total mass is also suppressed.
Meanwhile, lighter axion minihalos have a larger axion star formation rate due to the higher density, leading to a larger axion star to minihalo mass ratio $M_\star/M_h$.
The diagonal dashed blue line indicates the pivot minihalo mass, $x_{h,p}$ which is the $x_p$, or equivalently minihalo mass, that maximizes the axion star mass formed in the minihalo for a given axion mass.
The grey region presents the parameter space without any axion star formation around matter-radiation equality, which happens for larger axion minihalos or heavier axion masses.
An important note is that we focus on the white-noise part of the axion power spectrum, which allows us to derive all the halo properties analytically. However, axion minihalos with $x_h<1$ are predicted by simulations of axion string dynamics, which can potentially enhance the axion star formation rate \cite{Gorghetto:2024vnp}. Note that such axion minihalos usually do not contribute a significant mass fraction of the total axion dark matter population, which is dominated by the axion minihalos at the high mass end \cite{Xiao:2021nkb}. Therefore, as mentioned earlier, we focus on the white-noise part of the power spectrum which captures the most important physics in a simple analytic treatment.

The mass function of QCD axion stars is plotted in Fig.~\ref{fig:massfunc} for various axion masses. We labeled two contributions to the axion star mass function separately with solid and dashed curves, which come from lighter axion minihalos ($x_h<x_{h,p}$) and heavier axion minihalos ($x_h>x_{h,p}$) respectively. The actual mass function is the sum of these, but the lighter axion minihalo part always dominates, if it exists. Note that there is no solid line for $m_a=10^{-6}~ \eV$ because $x_{h,p}<1$ for this mass as shown in \Fig{fig:MRcontour}.
At the pivot minihalo mass, the mass function diverges because $dM_h/dM_\star$ diverges, as can be seen from the shape of the black $M_\star$ contours in \Fig{fig:MRcontour}. However, physical observables depend on the integration of the mass function, which is finite.
As we see in the plot, the total fraction of axion dark matter in axion stars, $f_\star$, monotonically decreases with axion mass. This behavior is related to the temperature-dependent mass of the QCD axion. The axion minihalo mass is determined by the critical time when axions start to oscillate and behave as nonrelativistic particles at $m_a(T)\approx H$. Usually, the QCD axion oscillates before the QCD epoch, where the axion mass is smaller than its zero temperature value after the QCD phase transition. This relatively late oscillation leads to a larger axion minihalo mass and a smaller axion star formation rate. As we see in Fig.~\ref{fig:massfunc_ALP}, the mass fraction in ALP stars is invariant with different axion masses since they start the oscillation at zero-temperature mass. As we discussed before, the mass function of axion stars is typically peaked at heavier masses as shown in Fig.~\ref{fig:massfunc}.

\subsection{ALP Stars}\label{sec:alpstars}

\begin{figure}[t]
\centering
\includegraphics[width=0.8\textwidth]{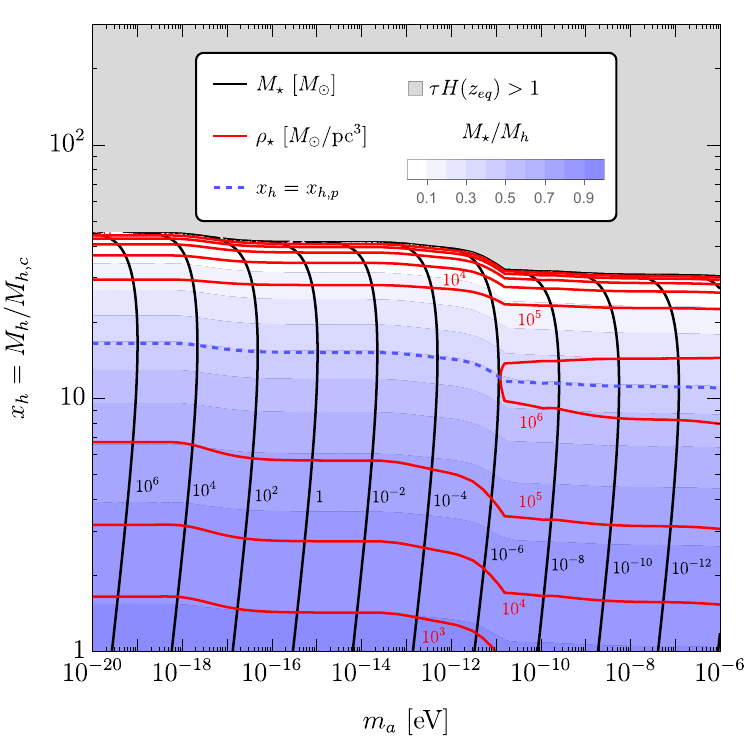} 
\caption{This figure has the same information as in Fig.~\ref{fig:MRcontour} but focuses on axion-like particles here. Again there is no axion star formation in the grey region because axion minihalos with lower masses have a larger density and thus a higher axion star formation rate. The characteristic axion minihalo mass $M_{h,c}$ is determined by the mass contained within the horizon when $m_a=H$, which is the canonical scenario for the vacuum misalignment mechanism. The red curves are the contours of axion star densities, which are nearly horizontal up to changes on $g_\star$. 
Note that the critical density of the Universe today is about $10^{-7}M_{\odot}/\rm pc^3$ while the axion stars have densities larger than $10^{3}M_{\odot}/\rm pc^3$. 
The black curves are the contours of axion star masses, which are approximately vertical and mostly determined by axion masses. A smaller $m_a$, corresponding to a longer de Broglie wavelength, will give a larger axion star mass. Therefore, axion stars with large masses and high densities can lead to interesting observational constraints.}
\label{fig:MRcontour_ALP}
\end{figure}

\begin{figure}[t]
\centering
    \includegraphics[width=0.95\textwidth]{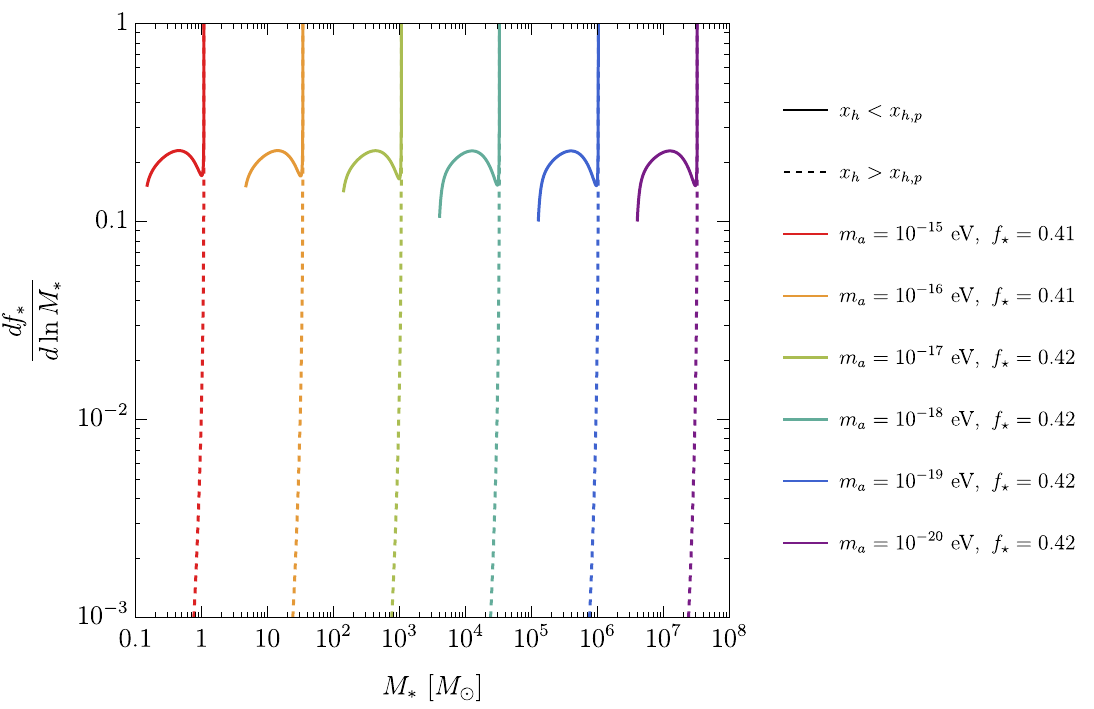}    
\caption{The mass function of ALP stars with the same scheme in Fig.~\ref{fig:massfunc}. The mass function is self-similar across different axion masses. This self-similarity differs from that in QCD axion stars where the axion mass is temperature-dependent around the initial axion oscillation time $m\sim H$. Still, the solid curves dominate the total mass of axion stars, suggesting that the low mass end of axion minihalos contributes most to the axion star population. A large fraction of axion dark matter, $f_\star \gtrsim 0.4$, is in the form of axion stars even though we only include the formation around matter-radiation equality.
}
\label{fig:massfunc_ALP}
\end{figure}

In this subsection, we consider an ALP that starts to oscillate at its zero temperature mass.  This determines the oscillation temperature through \cite{Blinov:2019rhb}
\begin{equation}\label{eq:Tosc}
    m_a = q_0 H(T_{\rm osc})= q_0\sqrt{\frac{\pi^2}{90}}\frac{g_*^{1/2}(T_{\rm osc}) T_{\rm osc}^2}{M_{\rm pl}},
\end{equation}
with $q_0=1.6$ and $T_{\rm osc}$ is the temperature of the Universe at the onset of axion oscillation. Note that we assume a standard cosmology where the Universe was dominated by radiation at early times, one could also consider alternative cosmological scenarios such as early matter domination \cite{Nelson:2018via,Blinov:2019rhb}. Since entropy is conserved after axion oscillation, we can relate the scale factor at oscillation to the current temperature via $g_{*S}(T_{\rm osc})a_{\rm osc}^3T_{\rm osc}^3=g_{*S}(T_0)T_0^3$, where $T_0$ is the current CMB temperature. Therefore we can obtain the follow expression for $k_{\rm osc}$
\begin{equation}
    k_{\rm osc} = \left(\frac{g_{*S}(T_0)}{g_{*S}(T_{\rm osc})}\right)^{1/3}\frac{T_0}{T_{\rm osc}}\frac{m_a}{q_0}.
\end{equation}
With this value of $k_{\text{osc}}$, and recalling that for an ALP $k_0=k_{\text{osc}}$ and $A_0\approx 0.3$, we can now repeat the process in the previous subsection to calculate $\Mhc$, $\Msc$, and $\tau$ for an ALP.  This results in
\begin{eqnarray}
\Mhc &\approx& 3.1 \, M_\odot \, \eta \left(\frac{10^{-16}~\eV}{m_a} \right)^{3/2}~, \\
\Msc &\approx& 22 \, M_\odot \, x_h^{1/12} \eta^{1/3} \left(\frac{10^{-16}~\eV}{m_a} \right)^{3/2}~, \\
\tau H(z_{\rm eq}) &\approx& \frac{1.6 \times 10^{-6} \, x_h^{7/2} \eta^2}{\log \left[0.78 \, x_h^{11/12} \eta^{2/3} \right]}~. \label{eq:tauALP}
\end{eqnarray}
Here, $\eta \equiv g_{*S}(T_\text{osc})/g_*^{3/4}(T_\text{osc})$ and $T_{\rm osc}$ is determined by $m_a$ from \Eq{eq:Tosc}. Note that the log factor in \Eq{eq:tauALP} is always positive. With the above expressions, we obtained all the equations needed to compute the axion star mass function for ALPs. Note that $\tau$ is independent of $m_a$ except for a minor dependence from $g_*$. This gives an almost universal mass distribution shape for all $m_a$, although as $m_a$ is increased the peak in the distribution of star masses also increases.

In Fig.~\ref{fig:MRcontour_ALP}, we present the same axion star properties as Fig.~\ref{fig:MRcontour} for axion-like particles. The black and red curves are the mass and density contours of ALP stars, respectively. We also show the pivot axion minihalo mass $x_{h,p}$ with the blue dashed line.
Similar to the QCD axion scenario, axion minihalos with smaller masses still have a larger axion star formation rate. Therefore smaller $x_h$ can lead to axion star formation while large $x_h$ correspond to the grey region without any axion stars forming during matter-radiation equality. 
Unlike QCD axion stars, the ALP star formation rate is independent of $m_a$ except for the minor $g_*$ dependence. Thus, the red curves and the boundary of the grey region are nearly horizontal. The mass contours are almost vertical because the de Broglie wavelength of axion waves sets the size of axion stars while the axion star densities are approximately independent of  $m_a$. A smaller $m_a$ corresponds to a larger $M_\star$, which implies constraints on massive compact objects can place a lower bound on the ALP mass. 

As shown in Fig.~\ref{fig:massfunc_ALP}, the shape of the mass function of ALP stars is invariant for different $m_a$. The shape slightly changes between $m_a \lesssim 10^{-17}~\eV$ and $m_a \gtrsim 10^{-17}~\eV$ due to the sudden change of $g_*$ from the electron-positron annihilations, but the total mass fraction does not change much. 
Note that each mass function has a divergent peak at $x_{h,p}$ as discussed in Sec.~\ref{sec:QCDaxionstars}.
As shown in Fig.~\ref{fig:massfunc_ALP}, the mass fraction of axions that have formed axion stars is $f_\star \sim 0.42$. Such a large axion star fraction can lead to interesting observational constraints such as dynamical heating effects in ultrafaint dwarfs \cite{Graham:2023unf,Graham:2024hah}, which we discuss in the next section. 

The axion star mass function shown in Fig.~\ref{fig:massfunc_ALP} shows that the distribution of axion star masses span roughly one order of magnitude, for a given axion mass. We have thus far only considered the formation process around matter-radiation equality. 
We expect that this dominates the overall rate because the axion star formation rate is peaked around this time due to the large density of axion minihalos formed at equality.  At later times the halos, as the halo masses grow, become more dilute, and the star formation time increases. We have ignored the merger of axion stars at later times, which may increase the average star mass and extend the star mass function at the high mass end.

\section{Constraints}\label{sec:observation}

\begin{figure}
\centering
    \includegraphics[width=0.8\textwidth]{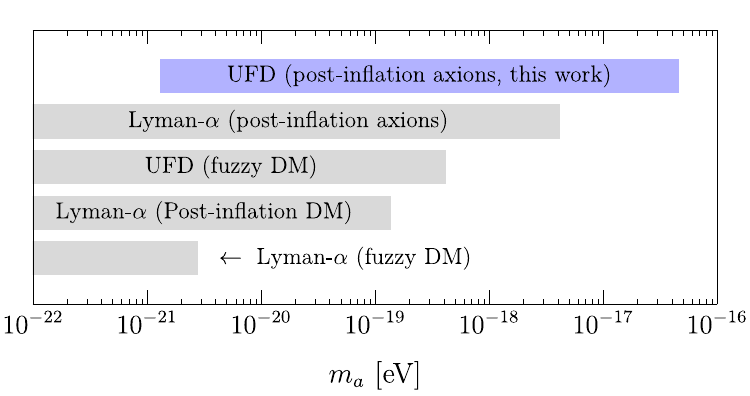}    
\caption{Current status of constraints on ultralight dark matter. Here we compare our results with the existing constraints in the literature that apply to various dark matter models. For generic fuzzy dark matter, there are Lyman-$\alpha$ constraints ($m_a> 2\times 10^{-21}$ eV \cite{Irsic:2017yje}) and ultrafaint dwarf (UFD) galaxies constraints ($m_a> 3\times 10^{-19}$ eV \cite{Dalal:2022rmp}). Lyman-$\alpha$ also constrains post-inflationary axions ($m_a> 3\times 10^{-18}$ eV \cite{Irsic:2019iff}), which is the same scenario as discussed here, but the bound in this work is stronger by one-order-of-magnitude ($m_a>3.3\times 10^{-17}\rm eV$). There is also a generic bound on post-inflation dark matter ($m_a>10^{-19}$ eV) considering the axion isocurvature perturbations and the free-streaming effect, which will be again constrained by Lyman-$\alpha$ \cite{Amin:2022nlh}. The constraint from ALP stars is derived by considering their dynamical heating effects in ultrafaint dwarfs \cite{Graham:2023unf,Graham:2024hah}.
}
\label{fig:constraints}
\end{figure}

Using the axion star mass functions determined above, we now turn to possible constraints on axion dark matter coming from the effects of these exotic compact objects.  As mentioned earlier, we focus only on axion stars interacting through gravitational interactions, to be as model independent as possible.  For ALPs, we considered a wider, and generally smaller, range of axion mass than for QCD axions which leads to axion stars which are more massive than in the case of the QCD axion.  We did consider axions of both types in the mass range $10^{-8}~\eV<m_a<10^{-6}~\eV$, and here although the range of axion star masses are comparable their distribution and the total mass fraction are different.  

Generally, over the QCD axion mass range the star masses are very light, $10^{-14} M_\odot \lesssim M_\star \lesssim 10^{-11} M_\odot$.  This means that QCD axion stars are not massive or dense enough to lead to meaningful constraints from galactic microlensing \cite{Croon:2020ouk}.  For the ALP scenario the star mass is much heavier, for the range of axions we considered we find stars in the range $0.1 M_\odot \lesssim M_\star \lesssim 10^{8} M_\odot$.  These masses are beyond the reach of microlensing searches, due to the low number density of the axion stars.  However, this mass range can instead be constrained by observations of ultra faint dwarfs, which we now discuss.

It has recently been pointed out that the motion of pointlike \cite{10.1093/mnras/sty079,zoutendijk:hal-02475222,Stegmann:2019wyz,Brandt:2016aco,Graham:2023unf} and diffuse dark matter substructures \cite{Graham:2024hah} inside dwarf galaxies can lead to heating of the stars in the dwarf, which in turn can place a constraint on the number of these structures.  The ensemble of dark matter substructures is typically hotter than the gas of stars in these systems and thus, through gravitational interactions, heat will flow from one system to the other.  This heating will increase the average speed, and thus the scale radius, of the stars in the dwarf.   From observations of the distribution of stars in Segue-I, an ultra-faint dwarf (UFD) galaxy, Refs.~\cite{Graham:2023unf,Graham:2024hah} placed strong constraints on the fraction of dark matter that can be in these substructures, as a function of the clump mass and density.  The UFD constraints were derived assuming that all clumps had the same mass and for each mass they can be interpreted as a bound on the fraction of dark matter that can be in substructures of that mass.  

For a given axion model our analysis gives a prediction for the axion star mass function, shown in Figs.~\ref{fig:massfunc_ALP}.  
Since the contribution of each substructure to heating UFD is linearly adding up, we can translate the constraints on a single mass distribution to the constraints on a specific mass distribution with
\be \label{eq:constraint}
 \int f_{\textrm{UFD}}^{-1}(M_\star,\rho_\star) \frac{d f_\star}{d M_\star} dM_\star \le 1\, ,
\ee
where $f_{\textrm{UFD}}$ is the constraint on the mass fraction of compact objects assuming a single mass $M_\star$ and density $\rho_\star$ in Ref.~\cite{Graham:2024hah}. Any axion star mass function that does not satisfy Eq.~\ref{eq:constraint} will be ruled out. We can easily test if certain ALP masses are excluded by applying the axion star mass function presented in Fig.~\ref{fig:massfunc_ALP} to the above integral Eq.~\ref{eq:constraint}. Lighter axion masses are more constrained because there are stronger bounds for more massive objects.
As a result, we exclude the ALP mass $1.8 \times 10^{-21}~{\rm eV}<m_a< 3.3 \times 10^{-17}~{\rm eV}$. We present our result in \Fig{fig:constraints} in comparison to other constraints on fuzzy DM, post-inflation DM, and post-inflation axions. Our work focuses on the observational constraints of post-inflation axions, which feature an enhancement on the matter power spectrum as well as the formation of axion minihalos and axion stars. The constraint on axion minihalos only from Lyman-$\alpha$ has already been considered and leads to a weaker constraint \cite{Irsic:2019iff}. There are also generic bounds on the mass of post-inflation dark matter \cite{Amin:2022nlh} and fuzzy dark matter \cite{Irsic:2017yje,Schutz:2020jox,Dalal:2022rmp,Zimmermann:2024xvd}, which are also weaker than our bound obtained from the gravitational effects of axion stars.

\section{Discussion}\label{sec:discussion}

In the post-inflationary axion scenario, where the PQ symmetry is broken after inflation has ended, the axion field has a white-noise spectrum of isocurvature perturbations at short scales.  Once structure can grow, after matter-radiation equality, these perturbations can quickly form axion minihalos.  The mass distribution of these minihalos can be calculated based on the spectrum of perturbations.  Furthermore, numerical simulations, and analytic insights, indicate that axion stars, compact bound states of axions held together by gravity, will form in the center of these minihalos.  

Using the growth rate of the stars inside minihalos and the mass function of the minihalos we determine the axion star mass function, $df_\star/dM_\star$, for both QCD axion and axion-like particles.  These results are useful for understanding the rate of possible signals coming from axion stars, such as the radio signals of axion stars, due to an axion-photon coupling, from their collisions with other objects.  We used the star mass function to determine the signal coming from the axion's gravitational interactions \emph{alone}.  In particular, the effect of axion stars on the heating of normal stars in ultra faint dwarfs.

On the dilute branch axion stars are stabilized by kinetic pressure and gravity, the mass-radius relation is simply determined by the axion mass. Therefore, the axion star mass function also contains information about the density and radius of axion stars.  We used this to translate recent bounds on heating in ultra faint dwarfs, by dark compact objects of a single mass, into bounds on axions.  For QCD axion stars, we did not find any current constraints. ALP stars, on the other hand, can be much more massive and this approach places a constraint of $m_a>3.3\times 10^{-17}$ eV, see Fig.~\ref{fig:constraints} for a comparison of constraints. In addition, the ALP star mass function suggests that almost half of the axion dark matter is in the form of ALP stars for a broad range of ALP masses.

There are several ways in which our approach can be improved.  First, our calculation is based on a pure white-noise power spectrum of axion isocurvature perturbations but axion strings will contribute to the power spectrum at higher wavenumber, altering the spectrum.  We did not look at minihalos outside of the white noise regime but including them would lead to more light minihalos.  For the ALP case, this enhancement is not very relevant because the lighter axion minihalos are not a significant fraction of the total axion dark matter mass. For QCD axions with $m_a\gtrsim 10^{-6}$ eV, this enhancement may be more important. Second, we focused on the axion star population formed around the time of matter-radiation equality, which is motivated by the fact that the axion star formation rate is on average peaked at earlier times due to the high density of the average minihalos. We have not considered late evolution such as the merger of axion stars, which leads to more massive and thus more detectable axion stars.  However, to include late time evolution one has to be careful not to double count mass in axion stars, a complete analysis would require careful tracking of stars in minihalos. We note that tidal disruption is only a problem for minihalos close to galactic centers and is unlikely to be an issue for denser axion stars.

\paragraph{Note added:} During the preparation of this work, Ref.~\cite{Gorghetto:2024vnp} appeared, studying the enhancements in QCD axion star formation based on the power spectrum from string simulations, which extends to smaller scales.  This allows the consideration of larger QCD axion masses that can form axion stars.  
The range of QCD axion masses considered is complementary to those considered here.

\section*{Acknowledgement}
We thank Peter Graham and Harikrishnan Ramani for the helpful discussions and for providing us with the data associated with their UFD constraint. We also thank Han Gil Choi, Wayne Hu, Xuheng Luo, and Dennis Maseizik for the useful discussions and Raymond Co for pointing out typos in some equations.
JHC, PJF, and HX are supported by Fermi Research Alliance, LLC under Contract DE-AC02-07CH11359 with the U.S. Department of Energy.
\bibliographystyle{jhep}
\bibliography{axion_star}

\end{document}